\documentclass{article}
\usepackage{spconf,amsmath,graphicx}

\usepackage{url,amsfonts,amssymb,bm}

\usepackage{enumitem}
\usepackage{algorithm}
\usepackage{algorithmicx, algpseudocode }
\usepackage{multirow}
\usepackage{color}
\usepackage{mathtools}
\usepackage{caption}
\usepackage{subcaption}
\usepackage{amsthm}

\usepackage[colorlinks=true, linkcolor=blue, citecolor=blue, urlcolor=blue]{hyperref}
\hypersetup{pdfcreator  = {LaTeX with hyperref package},
               pdfproducer = {dvips + ps2pdf} }

\usepackage{tikz}

\usetikzlibrary{calc}

\usepackage{tikz,spconf,amsmath,graphicx,bbm}

\usetikzlibrary{positioning}

\theoremstyle{definition}

\DeclareMathOperator*{\diag}{diag}

\usepackage{bm}

\long\def\comment#1{}

\newfont{\bbb}{msbm10 scaled 700}

\newfont{\bb}{msbm10 scaled 1100}

\newcommand{\av}{{\bf a}}
\newcommand{\bv}{{\bf b}}

\newcommand{\dv}{{\bf d}}

\newcommand{\vv}{{\bf v}}

\newcommand{\Dm}{{\bf D}}

\newcommand{\Qm}{{\bf Q}}

\newcommand{\Tm}{{\bf T}}

\newcommand{\Wm}{{\bf W}}
\newcommand{\Vm}{{\bf V}}

\newcommand{\Cc}{{\cal C}}

\newcommand{\Ec}{{\cal E}}

\newcommand{\Gc}{{\cal G}}

\newcommand{\Ic}{{\cal I}}

\newcommand{\Oc}{{\cal O}}
\newcommand{\Pc}{{\cal P}}
\newcommand{\Qc}{{\cal Q}}

\newcommand{\Deltam}{\hbox{\boldmath$\Delta$}}

\newcommand{\Phim}{\hbox{\boldmath$\Phi$}}

\title{ Multi-resolution intra-predictive coding of 3D point cloud attributes }
\name{Eduardo Pavez$^\star$,~Andr\'e L. Souto$^\#$,~Ricardo L. De Queiroz$^\#$,~and Antonio Ortega$^\star$ \thanks{This work was funded in part by NSE under grant  CNS-1956190. Author's email: pavezcar@usc.edu, andre@image.unb.br, queiroz@ieee.org, ortega@sipi.usc.edu.   }}
\address{$^\star$University of Southern California, Los Angeles, California, USA \\
$^\#$ Universidade de Brasilia, Brasilia, Brazil }
\begin{document}
\ninept
\maketitle
\begin{abstract}
We propose an intra frame predictive strategy for compression of 3D point cloud attributes.  Our approach is integrated with the region adaptive graph Fourier transform (RAGFT), a multi-resolution transform formed by a composition of localized block transforms, 
which produces a set of low pass (approximation) and high pass (detail) coefficients at multiple resolutions. Since the transform operations are spatially localized, RAGFT coefficients at a given resolution may still be correlated. To exploit this phenomenon, we propose an intra-prediction strategy, in which decoded approximation coefficients are used to predict uncoded detail coefficients. The prediction residuals are then quantized and entropy coded. For the 8i dataset, we obtain gains up to $0.5$db as compared to intra predicted point cloud compresion based on the region adaptive Haar transform (RAHT). 
\end{abstract}
\begin{keywords}
3D point clouds,  intra prediction, multiresolution transform, graph fourier transform
\end{keywords}
\section{Introduction}

 3D point clouds (3DPC) have become the preferred representation of  3D scenes, people and objects \cite{schwarz2018emerging,graziosi2020overview,pavez2018dynamic}. They  consist of a list of points coordinates in 3D space along with color  attributes.  Recent advancements in real time 3D capture, along with  potential applications to entertainment and immersive communications, have prompted research on 3DPC compression \cite{schwarz2018emerging,graziosi2020overview}.  
 
This paper focuses on 3DPC attribute compression.
Earlier approaches for compression of 3DPC attributes were based on  transform coding techniques, that is, transformation followed by quantization and entropy coding, similar  to modern image codecs. For 3DPCs, a popular approach is based on the region adaptive hierarchical (or Haar) transform (RAHT) \cite{de2016compression,GPCC:20}. In  image and video coding, intra and inter prediction are often combined with transform coding.  Predictive methods for 3DPC compression have only recently become popular, which may be explained by the fact that good spatial and temporal predictors are harder to obtain for 3DPCs, since these  
 i) represent complex surface and  non-surface geometries, ii) have spatial frame to frame irregularity, and iii) lack temporal consistency.
There has been substantial recent work to apply transform coding to inter frame (temporal) prediction residuals \cite{pavez2018dynamic,xu2020predictive,souto2020predictive,de2017motion,thanou2016graph}. However, intra frame prediction is less explored. In  \cite{cohen2016point,shao2018hybrid}, block based intra-prediction similar to video was considered. More recently, MPEG adopted an intra prediction strategy for the RAHT (I-RAHT) \cite{uraht:19}, which uses a multi-resolution prediction, instead of traditional directional block based prediction. 

\begin{figure}[t]
    \centering
   \begin{subfigure}{.23\textwidth}
  \centering
  \includegraphics[width=1\textwidth]{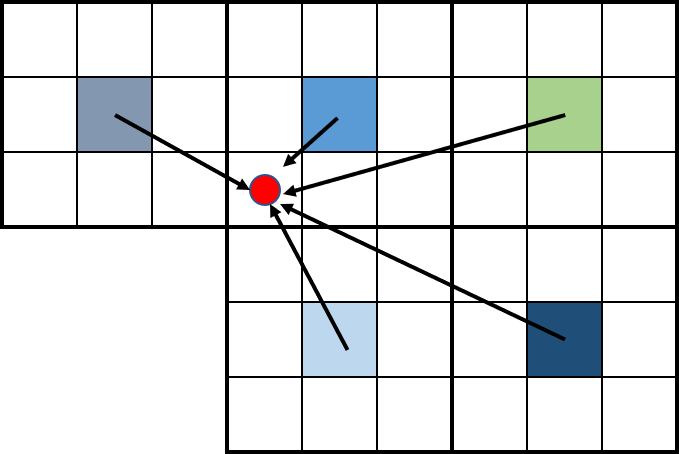}
  \caption{I-RAHT}
  \label{fig:smoothing_uraht}
\end{subfigure}
\begin{subfigure}{.23\textwidth}
  \centering
  \includegraphics[width=1\textwidth]{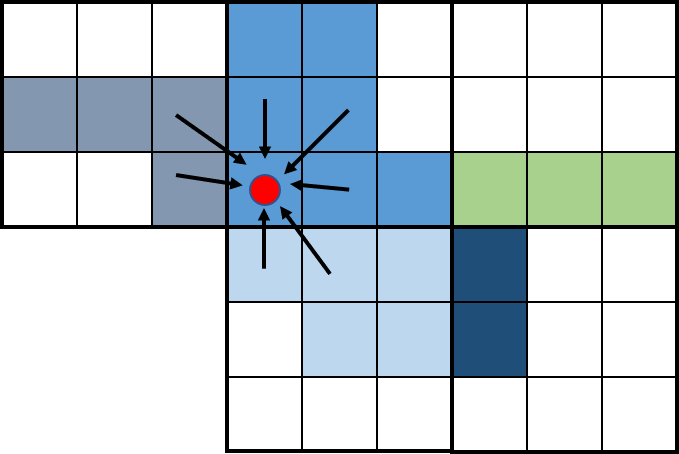}
  \caption{Proposed}
  \label{fig:smoothing_proposed}
\end{subfigure}
    \caption{Comparison of multi-resolution predictors. (\ref{fig:smoothing_uraht}) Predictor attribute (red dot) is a linear combination of neighbors in lower resolution point cloud ($\Vm_{\ell}$). (\ref{fig:smoothing_proposed}) Predictor attribute is a linear combination of neighbors in current resolution point cloud ($\Vm_{\ell +1}$).}
    \label{fig:smoothing}
\end{figure}

In  I-RAHT, low pass decoded RAHT coefficients are used to predict high pass RAHT coefficients. To achieve this, a higher resolution predictor point cloud is constructed from lower resolution decoded coefficients. Each predictor attribute in the high resolution point cloud is obtained as a linear combination of the nearest attributes in the lower resolution point cloud. Then the detail coefficients of this predictor signal are used to predict the target detail coefficients. While this approach to construct a higher resolution point cloud is suitable for the RAHT, since it is formed by a composition of $2 \times 2$ orthogonal transforms, it fails to   take into account the more complex geometry of the higher resolution point cloud. This becomes even more important  when considering larger block transforms
with more points used by the region adaptive graph Fourier transform (RAGFT) \cite{pavez2020ragft}, as shown in Figure \ref{fig:smoothing}.

 In this paper we propose an intra-predictive coding framework for  RAGFT (I-RAGFT). We use the same multi-resolution prediction strategy  used in I-RAHT, where decoded approximation coefficients are used to predict and code detail coefficients, as depicted in Figure \ref{fig:intra_coding}. 
 However, different from the prediction algorithm used by I-RAHT, our proposed predictors exploits  the higher resolution point cloud geometry. We start by projecting the low resolution signal onto the higher resolution geometry by zero padding and applying a one level inverse RAGFT. The resulting interpolated signal is piece-wise constant, as depicted by the colored voxels in Figure \ref{fig:smoothing_proposed}. Then a smoothing graph filter \cite{ortega2018graph,milanfar2012tour,cheung2018graph,girault2018irregularity} is applied using a graph constructed on top of the higher resolution point cloud. While our approach is new for 3DPCs, similar ideas have been used to improve image coding by predicting wavelet coefficients with   learning based super resolution algorithms \cite{dimitriadis2020augmenting}.
 
The difference between our proposed approach and I-RAHT is illustrated in Figure \ref{fig:smoothing}. I-RAHT uses  a \textit{block-level}  predictor, since the approximation coefficients of the nearest neighboring blocks are used (Figure \ref{fig:smoothing_uraht}), while our approach uses a \textit{point-level} predictor, where fine resolution point values are interpolated, then filtered (across block boundaries), so that only  nearby points (instead of nearby blocks) are used to compute the predictor.  
%
%
We show through compression experiments that the proposed approach  can outperform I-RAHT, when using uniform quantization and adaptive RLGR entropy coding \cite{malvar2006adaptive}. 

The rest of the paper is organized as follows. In Section \ref{sec_predictive} we introduce the multi-resolution intra-prediction coding framework. The proposed predictors for the RAGFT are described in \ref{sec_filter}. We show numerical experiments and conclusions in Sections \ref{sec_exp} and \ref{sec_conclusion}, respectively. 
 %
%
\section{Multi resolution predictive coding}
\label{sec_predictive}
Consider a point cloud with point coordinates stored in the $N \times 3$ matrix  $\Vm = [\vv_i]$, and attributes $\av = [a_i]$. We will assume the attributes are processed with a multi-resolution transform, which at each resolution level takes an attribute vector $\av_{\ell+1}$, and produces approximation and detail coefficients at resolution $\ell$
\begin{equation}\label{eq_analysis_2chan}
\begin{bmatrix}
\av_{\ell} \\
\dv_{\ell}
\end{bmatrix}
= \Tm_{\ell} \av_{\ell +1}.
\end{equation}
The transform matrix $\Tm_{\ell}$ is an orthonormal matrix, thus $\Tm_{\ell}^{-1} = \Tm_{\ell}^{\top}$.
We assume that the original 3DPC attributes are stored at the highest resolution, $L$, so that  $\av_{L} = \av$. After applying (\ref{eq_analysis_2chan}) $L$ times, we obtain transform coefficients
\begin{equation}
\label{eq_transform_coeffs}
\begin{bmatrix}
\av_0^{\top},  \dv_0^{\top},  \dv_1^{\top},  \cdots,  \dv_{L-1}^{\top} 
\end{bmatrix}^{\top}.
\end{equation}
Several orthonormal transforms for 3DPC attributes can be described this way, including the block based graph Fourier transform \cite{zhang2014point}, RAHT \cite{de2016compression} and  RAGFT \cite{pavez2020ragft}. 
Since RAGFT is a composition of spatially localized block transforms, there may be additional redundancy between transformed coefficients, similar to what is observed for the RAHT \cite{uraht:19}. While  previous approaches code the  coefficients in (\ref{eq_transform_coeffs}) independently ignoring their dependencies across blocks, in this work we exploit them   to improve coding efficiency. 

\begin{figure}[t]
    \centering
    \includegraphics[width=0.5\textwidth]{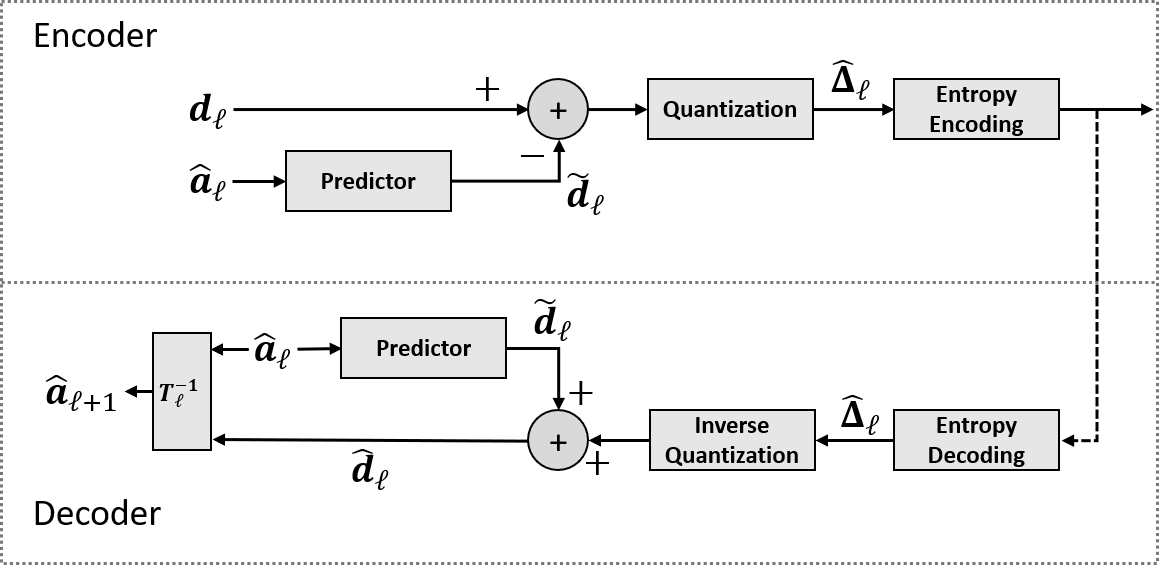}
    \caption{ Multi-resolution intra-predictive coding }
    \label{fig:intra_coding}
\end{figure}

Denote by   $\widehat{\Phim}_0 = \Qc(\av_o)$, the quantized low pass coefficients, where  $\Qc(\cdot)$ is a quantization operator, and  denote by $\hat{\av}_0 = \Qc^{-1}(\widehat{\Phim}_0 )$  the corresponding decoded coefficients. Now we define several quantities recursively. The decoded approximation coefficient at resolution $\ell$ is given by $\hat{\av}_{\ell}$, while the corresponding  detail coefficient at the same resolution is $\dv_{\ell}$. We will assume there is   function $\Pc_{\ell} (\hat{\av}_{\ell}) = \tilde{\dv}_{\ell}$, that predicts detail coefficients from approximation coefficients. This predictor will be explained in detail in the next section. Using $\Pc_{\ell}$ we compute a residual and  quantize it obtaining
\begin{equation}
    \widehat{\Deltam}_{\ell} = \Qc (\dv_{\ell} - \tilde{\dv}_{\ell}).
\end{equation}
If  the predictor is good enough,  coding  $\widehat{\Deltam}_{\ell}$ is more efficient  than coding the transform coefficients $\Qc (\dv_{\ell})$ directly.
The decoded details coefficients are given by
\begin{equation}
    \hat{\dv}_{\ell} = \tilde{\dv}_{\ell} + \Qc^{-1}(\widehat{\Deltam}_{\ell}).
\end{equation}
The decoded approximation and detail coefficients at resolution $\ell$ are used to obtain approximation coefficients at resolution $\ell+1$ with
\begin{equation}
    \hat{\av}_{\ell+1} = \Tm^{-1}_{\ell} \begin{bmatrix}
    \hat{\av}_{\ell} \\
    \hat{\dv}_{\ell} 
    \end{bmatrix}.
\end{equation}
The  quantities  $\widehat{\Deltam}_{\ell}$    can be computed recursively as depicted in Figure \ref{fig:intra_coding}, starting from the lowest  resolution coefficients $\av_0$.  
A typical transform coding strategy would quantize and entropy encode (\ref{eq_transform_coeffs}). In this work we encode 
\begin{equation}
\label{eq_residual_coeffs}
    \begin{bmatrix}
\widehat{\Phim}_0^{\top},  \widehat{\Deltam}_0^{\top},  \widehat{\Deltam}_1^{\top},  \cdots,  \widehat{\Deltam}_{L-1}^{\top} 
\end{bmatrix}^{\top}.
\end{equation}
\section{Multi resolution prediction}
\label{sec_filter}
In this section we describe the function  $\Pc_{\ell}(\hat{\av}_{\ell}) = \tilde{\dv_{\ell}}$, used  to predict the high pass coefficients $\dv_{\ell}$. 

\subsection{Graph representation of point clouds}
While forming transform coefficients (\ref{eq_transform_coeffs}), many  transforms \cite{de2016compression,pavez2020ragft}, either explicitely or implicitely,   produce  sets of point coordinates at various resolutions (e.g., by down-sampling), thus for each resolution $\ell$, we have  a point cloud $(\Vm_{\ell}, \av_{\ell})$, where $\Vm_L = \Vm$ and $\av_L = \av$.

For each of these point clouds   consider a graph $\Gc_{\ell} = (\Vm_{\ell}, \Wm_{\ell}, \Ec_{\ell})$. 
The matrix $\Wm_{\ell}$ is the adjacency matrix   and  $\Dm_{\ell} = \diag(\sum(\Wm_{\ell}))$ is the degree matrix. The graph has edge set $\Ec_{\ell}$, where $ij \in \Ec_{\ell}$ if point $\vv_{i,\ell}$ is ``near'' to point $\vv_{j,\ell}$.  Edge weights are given by
\begin{equation}
    (\Wm_{\ell})_{ij} = \frac{1}{\Vert \vv_{i,\ell} - \vv_{j,\ell} \Vert}.
\end{equation}
\subsection{Constructing the predictor}
\begin{figure}[ht]
    \centering
    \includegraphics[width=0.45\textwidth]{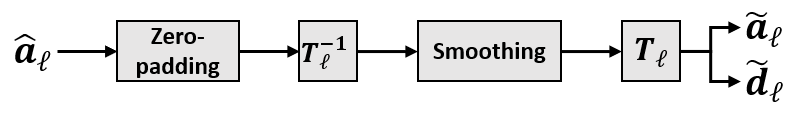}
    \caption{Proposed predictor $\tilde{\dv}_{\ell}=\Pc_{\ell}(\hat{\av}_{\ell})$. }
    \label{fig:predictor}
\end{figure}

The first step in constructing our predictor is interpolation of the low resolution point cloud $(\Vm_{\ell}, \hat{\av}_{\ell})$ by zero padding, and inverse transformation, leading to the  attribute signal %
\begin{equation}\label{eq_interpolation_zeropadding}
   \bv_{\ell+1} =  \Tm_{\ell}^{-1} \begin{bmatrix}
    \hat{\av}_{\ell} \\
    \mathbf{0}
    \end{bmatrix},
\end{equation}
and higher resolution point cloud $(\Vm_{\ell + 1}, \bv_{\ell+1})$. We propose applying  a smoothing (low pass) filter to the point cloud $(\Vm_{\ell + 1}, \bv_{\ell+1})$, and then applying the transform $\Tm_{\ell}$.  The \emph{predictor} block from Figure \ref{fig:intra_coding} is depicted in Figure \ref{fig:predictor}, with the caveat that  $\tilde{\av}_{\ell}$ is ignored. 
\begin{figure}[ht]
    \centering
    \begin{subfigure}{.5\textwidth}
  \centering
  \includegraphics[width=1\textwidth]{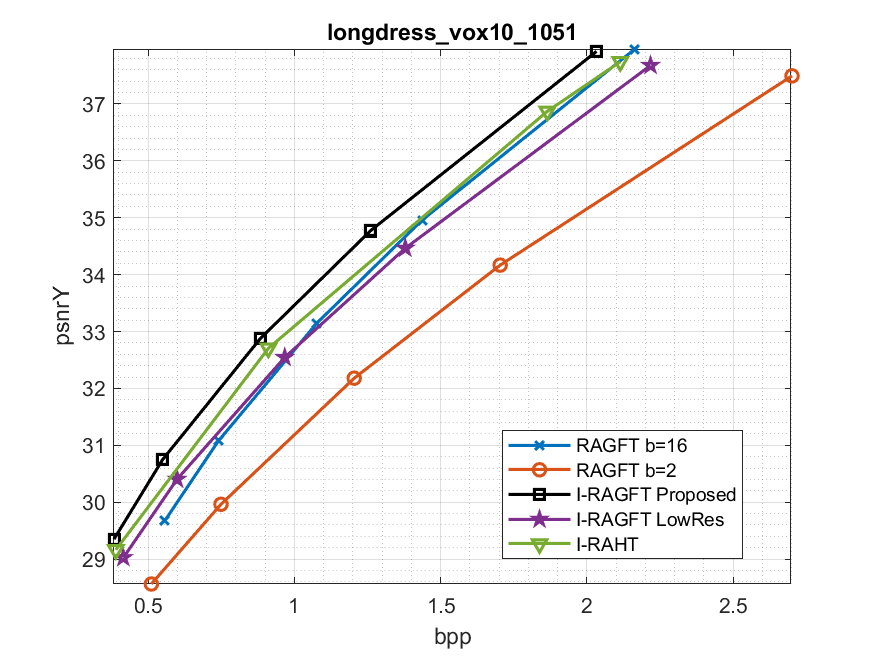}
  \caption{Longdress}
  \label{fig:rd_longdress}
\end{subfigure}
     \begin{subfigure}{.5\textwidth}
  \centering
  \includegraphics[width=1\textwidth]{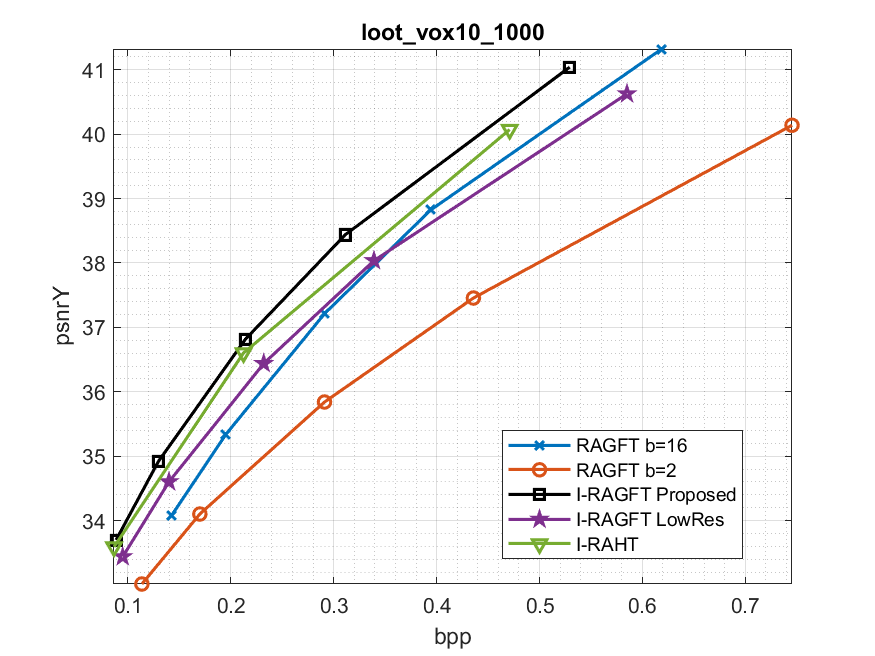}
  \caption{Loot}
  \label{fig:rd_loot}
\end{subfigure}
    \caption{Rate distortion curves for color compression. }
    \label{fig:rd}
\end{figure}
\begin{figure*}[ht]
    \centering
    \includegraphics[width =0.73 \textwidth]{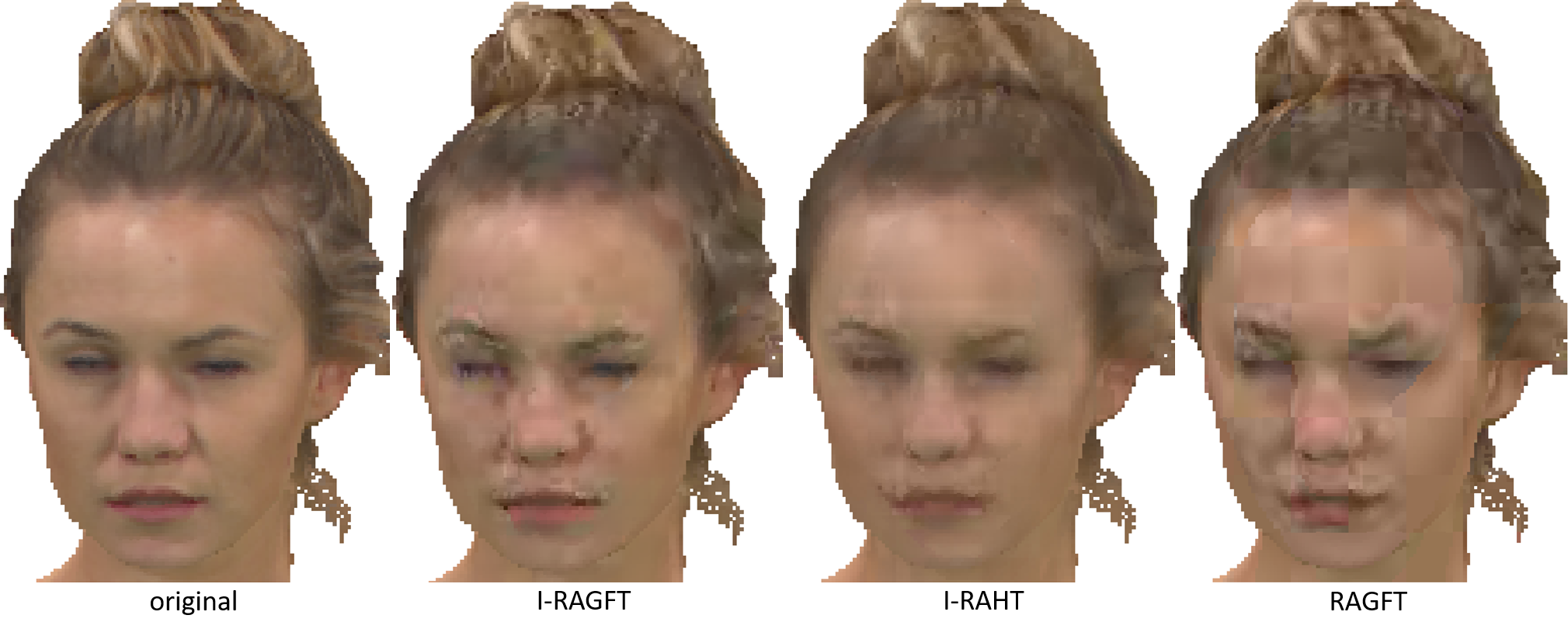}
    \caption{Comparison of reconstructed point clouds at  $0.493$ bits per point. I-RAGFT ($PSNR = 30.31 dB$), I-RAHT ($PSNR = 30 dB$), and RAGFT ($PSNR = 29.11 dB$) }
    \label{fig:longdress_head_subjective_comparison}
\end{figure*}
\subsection{Graph filtering of RAGFT coefficients}
In the RAGFT \cite{pavez2020ragft}, each transform  coefficient has an  importance weight that depends on the size (number of points) of the region of the point cloud they represent.  The $i$th point of the point cloud of   resolution $\ell$ is denoted by $\vv_{i,\ell}$. The point $\vv_{i,\ell}$ has a set of children at resolution $\ell +1$ denoted by $\Cc_{i,\ell}$. 
The index set of the children is denoted by $\Ic_{i,\ell} = \lbrace j: \vv_{j,\ell+1} \in \Cc_{i,\ell} \rbrace $. We denote the weight of point $\vv_{i,\ell}$  by $q_{i,\ell}$.
Based on this parent-child relation, the weights can be computed recursively as 
\begin{equation}
    q_{i,\ell} = \sum_{j \in \Ic_{i,\ell}} q_{j,\ell + 1},
\end{equation}
and the weights at full resolution are   $q_{i,L} = 1$ for all $i$.
For the RAGFT,   the interpolation equation (\ref{eq_interpolation_zeropadding}) has a closed form, thus the $j$th entry of $\bv_{\ell + 1}$ is equal to
\begin{equation}
    b_{j,\ell+1} = \sqrt{\frac{q_{j,\ell+1}}{q_{i,\ell}}} \hat{a}_{i,\ell},
\end{equation}
where point $\vv_{i,\ell}$ is the parent of $\vv_{j,\ell+1}$.
Note that if we define the diagonal matrix of weights by $\Qm_{\ell+1} $, with $jj$ entry equal to $q_{j,\ell+1}$, then the  interpolated signal is 
\begin{equation}\label{eq_interpolation_piecewise}
      \bv_{\ell+1}  = \Qm^{1/2}_{\ell+1} \left(\sum_{i} \frac{\hat{a}_{i,\ell}}{\sqrt{q_{i,\ell}}} \mathbf{1}_{\Ic_{i,\ell}}\right),
\end{equation}
where $\mathbf{1}_{\Ic_{i,\ell}}$  is the indicator of the set  $\Ic_{i,\ell}$. Thus after normalization by the square root of the point weights, the interpolated signal is piece-wise constant. In fact, this signal is constant within cube shaped regions, because the RAGFT is is a composition of localized ``block'' transforms. Figure \ref{fig:smoothing_proposed} depicts this piece-wise constant signal.  Given that  $\bv_{\ell+1}$ has this form, we   take into account the normalization matrix $\Qm_{\ell+1}$, and the piece-wise constant structure, when designing our smoothing operator. The first step is  to  normalize the entries of $\bv_{\ell+1}$ by their point weights using $\Qm^{-1/2}_{\ell+1}$. The resulting piece-wise constant signal is filtered to smooth out boundaries between regions. Finally,  we scale back each attribute using  the matrix $\Qm^{1/2}_{\ell+1}$. 
The proposed filtering operation has the   form 
\begin{equation}
  \tilde{\bv}_{\ell+1} = \Qm^{1/2}_{\ell+1} \Dm^{-1}_{\ell+1} \Wm_{\ell+1}   \Qm^{-1/2}_{\ell+1}  \bv_{\ell+1}.
\end{equation}
At resolution $\ell$, the adjacency matrix $\Wm_{\ell}$ is constructed using $k$ nearest neighbors on the point cloud geometry $\Vm_{\ell}$.
The  signal $\tilde{\bv}_{\ell+1}$ is  transformed using (\ref{eq_analysis_2chan}), resulting in $\tilde{\av}_{\ell}$ and $\tilde{\dv}_{\ell}$.
\subsection{Complexity}
The proposed intra predicted RAGFT can be computed by performing a forward and inverse RAGFT, along  with a  graph construction and a graph filtering operation per resolution level. 
For a point cloud with $N$ points the forward and inverse RAGFT have complexity $\Oc(N)$. If the octree \cite{jackins1980oct,meagher1982geometric} has to be computed, overall complexity increases to $\Oc(N\log(N))$. If the number of points at resolution $\ell$ is equal to $N_{\ell}$, computation of $\bv_{\ell}$ using (\ref{eq_interpolation_piecewise}) takes $\Oc(N_{\ell})$ time. Graph construction with $k$ nearest neighbors requires $\Oc(k N_{\ell}\log(N_{\ell}))$ operations, while $\tilde{\bv}_{\ell}$ can be computed using sparse matrix vector products in $\Oc(k N_{\ell})$ time. If $k$ is  $\Oc(1)$, then prediction operations at all resolutions have complexity $\Oc(N\log(N))$, resulting in an overall $\Oc(N\log(N))$ complexity of the intra predicted RAGFT. 
%
%
%
\section{Numerical results}
\label{sec_exp}
 We integrate  intra prediction with the RAGFT with $2 \times 2 \times 2$ blocks at all resolution levels, thus each localized block transform processes at most $8$ points. The proposed predictors are implemented using  graph filters, that use $K$ nearest neighbor graphs, with $K=7$ neighbors per point. We denote this approach by ``I-RAGFT''. We also implement predictors similar to those used by the G-PCC implementation of I-RAHT using KNN from the lower resolution graph (see Fig. \ref{fig:smoothing_uraht}). For this approach we set $K=5$, found after optimizing for best performance. This approach is denoted by ``I-RAGFT LowRes''.
We also compare against the RAGFT with high resolution block size equal to $16$, which achieves highest coding performance compared to  RAHT, and the RAGFT with high resolution block size equal to $2$, which provides a mild improvement over RAHT (see \cite{pavez2020ragft} for details). For  RAGFT and intra RAGFT, we uniformly quantize coefficients and entropy code them using  adaptive run-length Golomb-Rice (RLGR) algorithm \cite{malvar2006adaptive}.

The state of the art in non-video-based compression of  point cloud attributes,  called G-PCC \cite{GPCC:20} uses I-RAHT.  Several techniques are implemented in G-PCC to improve coding efficiency beyond that obtained through intra prediction.  Some of them include jointly encoding YUV coefficients  that are equal to zero, and adaptive quantization schemes of AC coefficients. These techniques could also be applied to the RAGFT but their implementation goes beyond the scope of this paper and we leave them for future work. 
In order to obtain a performance comparison between our method and the current state-of-the-art, under the same conditions,  we changed the source code of G-PCC modifying both the quantizer and the entropy encoder schemes. The adaptive quantizer was replaced with a uniform quantizer and the entropy encoder was modified to an adaptive RLGR algorithm. Moreover, we encode YUV coefficients independently. In Figure \ref{fig:rd} we report rate distortion curves for the ``longdress'' and ``loot'' point clouds of the  ``8iVFBv2'' dataset \cite{d20178i}.

Incorporating intra prediction into the RAGFT improves coding performance significantly. The difference between ``RAGFT $b=2$'' and ``I-RAGFT LowRes'' is about $1.5$db, while the gain obtained by using better predictors (``I-RAGFT Proposed'') can reach up to $2.5$db. 
In \cite{pavez2020ragft}, the RAGFT implemented with larger block sizes ($b_L=16$) led to the best results, since  spatial redundancy can be removed more efficiently  using graph transforms on larger blocks. Our results show that by  combining low resolution intra prediction and  RAGFT with small block size (``I-RAGFT $b=2$''), we    can outperform the best RAGFT at low bitrates. However this approach is still inferior to I-RAHT. The proposed  predictor based on the higher resolution point cloud,  combined with the RAGFT achieves the best rate distortion performance among all methods considered, and outperforms I-RAHT by up to $0.5$db. 

In Figure \ref{fig:longdress_head_subjective_comparison} we compare decoded attributes obtained after compression using I-RAGFT, I-RAHT and RAGFT ($b_L=16$), at the same rate of $0.493bpp$. Transform coding with the RAGFT produces strong blocking artifacts, similar to those associated to JPEG encoding. Both  intra coding approaches  have improved visual quality with respect to RAGFT, while also using smaller  transforms, thus avoiding blocking artifacts. I-RAHT produces an over smoothed reconstruction, which can be accredited to the less localized filtering operations. Since the  I-RAGFT uses higher resolution predictors,  image details are better preserved,  however it suffers from other localized artifacts. 
%
%
%
%
%
%
\section{Conclusion}
\label{sec_conclusion}
We studied the use of intra prediction for compression of 3D point cloud attributes with the RAGFT.  As with the RAHT, we showed that it is possible to remove redundancy between RAGFT  coefficients at different resolutions.  In the RAHT, a high resolution point cloud is predicted from a lower resolution point cloud. While this approach is efficient for predicting RAHT coefficients, it is less effective for the RAGFT. This is because the RAGFT uses larger block transforms, that makes prediction more challenging. To overcome this issue, we proposed a different prediction based on interpolation and graph signal filtering,  that takes into account the higher resolution geometry, thus  adapting better to the RAGFT. Compression experiments show the proposed approach  outperforms intra predictive coding of RAHT coefficients at all rates  by up to $0.5$db.
\bibliographystyle{IEEEbib}
\bibliography{refs}


\end{document}